\documentclass[twocolumn]{aa}
\usepackage{graphicx}
\usepackage{txfonts}
\usepackage[authoryear]{natbib}
\bibpunct{(}{)}{;}{a}{}{,}
\newcommand{\src} {4U\,2206+54}

\newcommand{\ha}  {H$\alpha$}

\def\simless{\mathbin{\lower 3pt\hbox
     {$\rlap{\raise 5pt\hbox{$\char'074$}}\mathchar"7218$}}}   
\def\simmore{\mathbin{\lower 3pt\hbox
     {$\rlap{\raise 5pt\hbox{$\char'076$}}\mathchar"7218$}}}   

\def\msun{~{\rm M}_\odot}

\begin{document}

   \title{Discovery of slow X-ray pulsations in the high-mass X-ray binary \src }

   \subtitle{}
  \author{
   P. Reig\inst{1,2}
          \and
	  J.M Torrej\'on\inst{3,4}
	  \and
	  I. Negueruela\inst{3}
	  \and
	  P. Blay\inst{5}
	  \and
	  M. Rib\'o\inst{6}
	  \and
	  J. Wilms\inst{7}
          }

\authorrunning{Reig et~al.}
\titlerunning{Slow X-ray pulsations in 4U 2206+54}

   \offprints{P. Reig}

   \institute{IESL, Foundation for Reseach and Technology-Hellas, 71110, 
   		Heraklion, Greece 
	 \and Physics Department, University of Crete, 71003, 
   		Heraklion, Greece 
		\email{pau@physics.uoc.gr}
        \and Departamento de F\'{\i}sica, Ingenier\'{\i}a de Sistemas y Teor\'{\i}a
de la Se\~nal, Universidad de Alicante, E-03080 Alicante, Spain
	\and Kavli Institute for Astrophysics and Space Research, Massachusetts Institute of
Technology, Cambridge MA 02139, USA
	\and Institut de Ciencia dels Materials, Universitat de Valencia, 
46071 Paterna-Valencia, Spain 
	\and Departament d'Astronomia i Meteorologia, Universitat de Barcelona,
Mart\'{\i} i Franqu\`es 1, 08028 Barcelona, Spain
	\and Dr. Karl Remeis-Observatory, University of Erlangen-Nuremberg, 
Sternwartstrasse 7, 96049 Bamberg, Germany 
	}

   \date{Received ; accepted}

\abstract
{The source \src\ is one of the most enigmatic high-mass X-ray binaries. In spite of 
intensive searches, X-ray pulsations have not been detected in the time
range $10^{-3}$-$10^{3}$ s. A cyclotron
line at $\sim$ 30 keV has been suggested by various authors but never
detected with significance. The stellar wind of the optical companion is 
abnormally slow. The orbital period, initially reported to be 9.6 days,
disappeared and a new periodicity of 19.25 days emerged.} 
{The main objective of our RXTE monitoring of \src\ is to study the 
X-ray orbital variability of the spectral and timing parameters. The new long 
and uninterrupted RXTE observations allow us to search for long ($\sim$1
hr) pulsations for the first time.  }
{We divided the $\sim$7-day observation into five intervals and obtained 
time-averaged energy spectra and power spectral density for each observation 
interval. We also searched for pulsations using various algorithms.}
{We have discovered 5560-s pulsations in the light curve of \src. Initially
detected in RXTE data, these pulsations are also present in
INTEGRAL and EXOSAT observations. The average 
X-ray luminosity in the energy 
range 2--10 keV is $1.5 \times 10^{35}$ erg s$^{-1}$ with a ratio 
$F_{\rm max}/F_{\rm min}\approx 5$. This ratio implies an eccentricity of
$\sim$0.4, somewhat higher than previously suggested. The power spectrum is 
dominated by red noise that can be fitted with a single power law whose index 
and strength decrease with X-ray flux. The source also shows 
a soft excess at low energies. If the soft excess is modelled with a blackbody 
component, then the size and temperature of the emitting region agrees with its 
interpretation in terms of a hot spot on the neutron star surface.
}
{The discovery of X-ray pulsations in \src\ confirms the neutron star nature of the
compact companion and definitively rules out the presence of a black hole. 
The source displays variability on time scales of days, presumably due to
changes in the mass accretion rate as the neutron star moves around the
optical companion in a moderately eccentric orbit. If current models for the
spin evolution in X-ray pulsars are correct, then the magnetic field of \src\ at
birth must have been $B\simmore 10^{14}$ G.}


   \maketitle

\begin{figure}
\resizebox{\hsize}{!}{\includegraphics{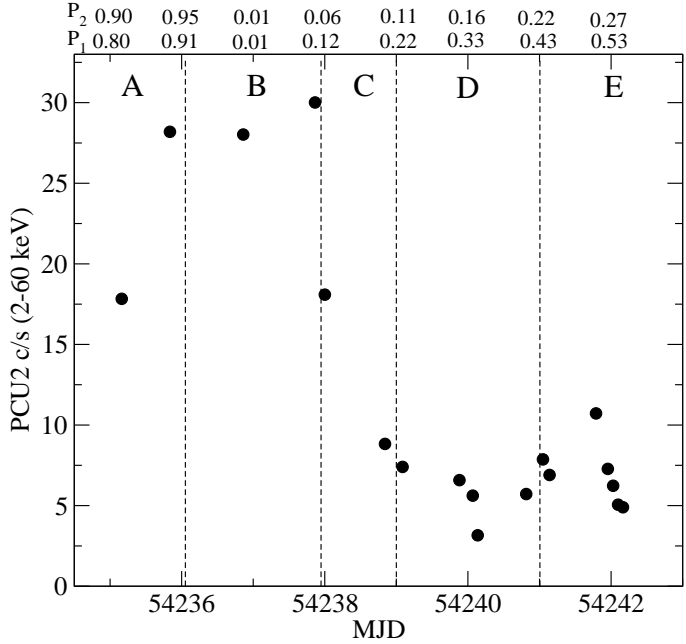}} \\
\caption[]{Light curve of \src\ covering the entire interval of the {\it
RXTE PCA}
observations. The data points represent the average of each observation
interval as given in Table~\ref{xobs}. Errors are less than the 
size of the points. Dashed lines separate the different
intervals for orbital variability analysis (see text). The numbers on top 
of the figure are the orbital phase for $P_1=9.56$ d and $P_2=19.12$ d. 
Reference time for phase calculation is 
MJD 53567.74, which corresponds to maximum flux in the folded ASM light
curve \citep{corb07}.}
\label{lc}
\end{figure}
\begin{table}
\begin{center}
\caption{Log of the RXTE observations.}
\label{xobs}
\begin{tabular}{@{~~}c@{~~}c@{~~}c@{~~}c@{~~}c}
\hline \hline \noalign{\smallskip}
ObsID	&MJD start--MJD stop	&On-source	&I$_{\rm X}^a$ &Obs. \\
	&	&time (s)	&c s$^{-1}$   &int. \\
\hline \noalign{\smallskip}
01-01-05 &54235.1628--54235.1908	&1600		&17.8 &A \\
01-01-02 &54235.8369--54236.1095 	&23472		&28.2 &A  \\
01-01-03 &54236.8613--54237.1036 	&20672		&28.0 &B  \\
01-01-04 &54237.8639--54238.0000 	&11696		&30.0 &B  \\
01-02-00 &54238.0010--54238.0841 	&7120		&18.1 &C  \\
01-02-01 &54238.8408--54239.0647 	&17328		&8.8 &C  \\
01-02-01 &54239.0865--54239.1199 	&2384		&7.4 &D  \\
01-02-02 &54239.8821--54240.0452 	&11328		&6.6 &D  \\
01-02-02 &54240.0671--54240.1124 	&2512		&5.6 &D  \\
01-02-02 &54240.1365--54240.1799 	&2256		&3.2 &D  \\
01-02-03 &54240.8137--54241.0263 	&13584		&5.7 &D  \\
01-02-03 &54241.0482--54241.0930 	&2528		&7.9 &E  \\
01-02-05 &54241.1412--54241.1604 	&400		&6.9 &E  \\
01-02-040 &54241.7899--54241.9452	&9328		&10.7 &E  \\
01-02-040 &54241.9545--54242.0069	&3776		&7.3 &E  \\
01-02-040 &54242.0287--54242.0743	&2544		&6.2 &E  \\
01-02-04 &54242.0982--54242.1408 	&2256		&5.1 &E  \\
01-02-04 &54242.1663--54242.1991 	&1984		&4.9 &E  \\
 \hline \noalign{\smallskip}
\multicolumn{5}{l}{$a$: Background subtracted 4--30 keV, PCU2 only}\\
\end{tabular}
\end{center}
\end{table}

\section{Introduction}

The X-ray source \src\ is one of the most enigmatic high-mass X-ray binaries (HMXB). Although
it has been studied by numerous ground- and space-based observatories, many
questions about the nature of its variability patterns remain unanswered.
In the optical and UV bands, the emitting spectrum is complex
\citep{negu01} and far from that of a Be star. Initially classified as a
B0-2 star \citep{stei84}, the optical counterpart defies any standard
spectral analysis. The absence of any obvious long-term trends in the
evolution of the \ha\ line spectral parameters and the
lack of any correlation between the \ha\ line equivalent width and infrared
magnitudes and colours  rules out a Be classification. However, the
strength and shape of the \ha\ line do not resemble those seen in
supergiant stars either. Although the
presence of an O9.5V star is virtually secured, the He abundance is
abnormally strong \citep{blay06}. 

Without a circumstellar disc around the donor, the material needed for
accretion and production of X-rays must come from the stellar wind.
However, the expected X-ray luminosity using the typical wind velocities of
an O9.5V star is about two orders of magnitude lower than the observed
X-ray luminosity $L_{\rm X}~\sim 10^{35}$ erg s$^{-1}$. The distance to
\src\ is estimated to be 2.6 kpc \citep{blay06}. \citet{ribo06} found that
the wind terminal velocity of \src\ is abnormally slow ($\sim$350 km
s$^{-1}$). With such a low velocity, an eccentric orbit and using the
Bondi-Hoyle formalism it is possible to reproduce the average X-ray
luminosity of the system as well as the orbital variability seen in the
{\it RXTE}/ASM light curve between 1996 and 2005. \src\ is the only
permanent wind-fed HMXB with a main-sequence donor \citep{ribo06}.

Equally enigmatic is the nature of the compact companion, which has not
been settled beyond doubt. Although a neutron star is the most likely
scenario, the low X-ray luminosity, compared to other persistent HMXB, has
been used as an argument to suggest the presence of a white dwarf
\citep{sara92,corb01}, while the lack of pulsations and the similarities
with LS 5039 do not allow us to rule out a  black hole
companion \citep{negu01} -- the most recent results show, however, that LS
5039 is probably a non-accreting neutron star \citep{ribo08}.

Broad-band X-ray observations favour the presence of a neutron star
\citep{torr04}. When modelled with Comptonization models, the resulting
seed photons have $kT>1$ keV, while the bolometric luminosity is low
($L_{\rm X}\approx 10^{35}$ erg s$^{-1}$). The only way to reconcile these
two results is by invoking a small emitting area ($< 2$ km), like that of a
polar cap in a neutron star \citep{mase04}, also ruling out the presence of
an accretion disc \citep{torr04}. The presence of a magnetic field with
$B\sim 2 \times 10^{12}$ G is implied by the possible detection of a
cyclotron resonance scattering feature at $\sim$30 keV. Although the
statistical significance of this feature is only marginal, it has been
observed by three different observatories, namely {\it RXTE}, {\it
BeppoSAX} and {\it INTEGRAL} \citep{torr04,mase04,blay05}. 


Another debated question is the orbital period of the system.
\citet{corb01} reported an orbital period of 9.6 days. This value was based
on observations made with the {\it RXTE} ASM instrument over 5 years. This
orbital period was refined to 9.5591 days by \citet{ribo06} using 9 years
worth of data of the same instrument. However, \citet{corb07} could not
find the 9.6-day modulation in the light curves of {\it SWIFT} BAT and {\it
RXTE} ASM for observations carried out from 2004 to 2005. Instead, they
found a strong modulation at 19.25 d, that is, consistent with twice
the 9.6-day period (or 19.12 d).

In this paper we investigate the X-ray variability of \src\ on time scales
of hours to days. The main objective is to search for $\sim$1-hr pulsations
and study possible spectral and timing orbital variability.

\begin{figure}
\resizebox{\hsize}{!}{\includegraphics{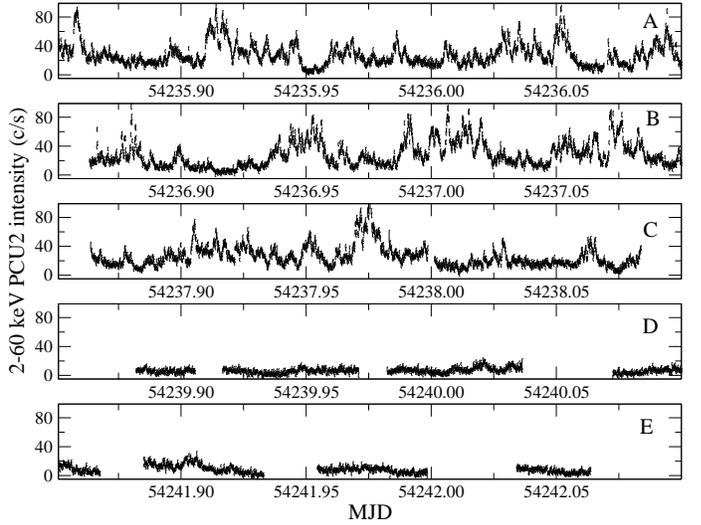}} \\
\caption[]{Portion of $\sim 6$ hours of the light curves of \src\ corresponding to each one
of the observation intervals. The axis scales were left the same
in all panels to allow easy comparison of the variability amplitude. The bin 
size is 10 seconds. Intervals C and D are more affected by observational
gaps. }
\label{lc_int}
\end{figure}
\begin{figure}
\resizebox{\hsize}{!}{\includegraphics{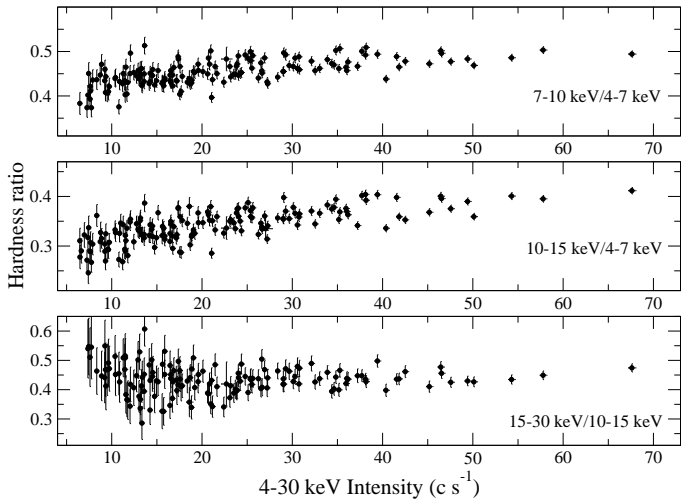}} \\
\caption[]{Hardness-intensity diagram of \src\ for various X-ray
bands. Count rates correspond to PCU2. Bin size is 512 s.}
\label{hr}
\end{figure}

\section{Observations}


\src\ was observed by the Rossi X-ray Timing Explorer ({\it RXTE}) during
the interval 15--22 May 2007  (JD 2454235.1628--2454242.1991) for a total
on-source time of 136.8 ks. Observation time with {\it RXTE} was
awarded in response to the A11 call for proposals and corresponds to the
{\it RXTE} proposal P92068. Light curves, spectra and response matrices
were extracted using the tools available in the {\it RXTE}
subpackage of version v.6.4 of the FTOOLS software package. 

{\it RXTE} is designed to facilitate the study of time variability in the
emission of X-ray sources with moderate spectral resolution. The pay-load
consists of a Proportional Counter Array (PCA), which is sensitive to
X-rays in the 2-60 keV energy range and has a total effective area of 6500
cm$^2$, shared by five PCU or proportional counter units 
\citep{jaho96}.
PCA data can be collected and telemetered to the ground in many different
ways depending on the intensity of the source and the spectral and timing
resolution desired. In this work we used the two standard modes: {\em
Standard1} provides 0.125-s resolution and no energy resolution; in the
{\em Standard2} configuration data are accumulated every 16 seconds in 129
channels. The High Energy X-ray Timing Experiment (HEXTE) comprises two
independent clusters of detectors with a total collecting area of 1400
cm$^2$, covering the energy range 15-250 keV. The spectral resolution is
$\sim$15\% at 60 keV energy range. During normal operation, the two
clusters rock on and off target to collect background data 
\citep{roth98}. Due to
malfunction of the rocking mechanism, cluster A was left permanently on the
on-source position for the entire duration of the observations. Thus, no
background light curve could be obtained. Data presented in this work
correspond to cluster B only. The third instrument on board {\it RXTE} is
the  All-Sky Monitor (ASM),which produce daily flux averages in the energy
range 1.3-12.1 keV at 30 mCrab sensitivity \citep{levi96}.  

The log of the {\it RXTE} observations is shown in Table~\ref{xobs}. Good time
intervals were defined  when the pointing of the satellite was stable
($<0.02^{\circ}$ from the source), the elevation above 8$^{\circ}$ and far
away from the South Atlantic Anomaly.  
Due to {\it RXTE}'s low-Earth orbit, the data consist of a number of
contiguous data intervals (typically 1 hr long) interspersed with
observational gaps produced by Earth occultations of the source and
passages of the satellite through the South Atlantic Anomaly.  In general,
the visibility of a target depends on its position with respect to the
plane of {\it RXTE}'s orbit around the Earth.  Objects that are within a
critical angle from the normal to the orbit will be visible for longer
periods of time.  In fact, if the object is close enough to the normal then
it would lie in the so-called Continuous Viewing Zone (CVZ).  Since {\it
RXTE} orbits relatively close to the Earth's equator, the regions of high
visibility and the CVZ are close to the north and south pole of the Earth. 
\src\ is at high ecliptic latitude (about 58 degrees), hence it has a
relatively high visibility. The observations reported here are considerably
longer than commonly reported for other sources. There are two long ($\sim$
6 hr) uninterrupted observations. These longer observations, together
with the monitoring of \src\ over 7 days, allow us to give new insights into
the properties of the X-ray emission in \src. In particular, they allow us
to search for periodicities on timescales of $\sim$ 1 hr. To investigate
potential orbital variability, the {\it RXTE} observations were divided
into five intervals with roughly the same on-source time, 
$\sim$20--30 ks (Fig.~\ref{lc}).

In addition to the {\it RXTE} pointing observations, we also analysed
archived data from the {\it INTEGRAL} and {\it EXOSAT} missions. The {\it
International Gamma-ray Astrophysics Laboratory} ({\it INTEGRAL})
observations were made between 16--19 December 2006 during revolution 510.
{\it INTEGRAL} carries two main gamma-ray instruments: the high angular
resolution Imager (IBIS) and the high-energy resolution spectrometer (SPI).
In our study we only used data from the upper detector system of IBIS,
called the Soft Gamma-ray Imager (ISGRI). ISGRI is sensitive to photons in
the energy range 20 keV to 1 MeV, has an angular resolution of 12' and a
source location accuracy that depends on the significance of the
detection: for a source at the limit of detection the source location
accuracy is about 4-5' while for a strong source it can go down to about
0.5' \citep{gros03}. The energy resolution is 7\% at 100 keV 
\citep{uber03}. \src\ was detected in a total of 19 {\it INTEGRAL} science
windows of around $\sim$2.3 ks exposure each. In all these observations the
satellite followed a $5\times5$ dithering pattern.  We only analysed images
in which the source fell in the fully coded field of view ($9^{\circ}
\times 9^{\circ}$). Data reduction was done with the Offline Scientific
Analysis (OSA) software version 7 \citep{cour03}.

The European Space Agency's X-ray Observatory, {\it EXOSAT}, was
operational from May 1983 to April 1986. The pay-load consisted of  2
Wolter Type I grazing incidence Low Energy (LE; 0.05-2 keV) Imaging
Telescopes, a Medium Energy (ME) Proportional Counter and a Gas
Scintillation (GS) Proportional Counter.  The ME detectors covered the
energy range 1-50 keV and had a field of view of 45 arcmin and an effective
area 1600 cm$^2$ \citep{turn81}. For our study we retrieved the ME detector standard
products available in the NASA's HEASARC archive. {\it EXOSAT} observed \src\
with the Medium Energy (ME) proportional counter in three occasions: 8
August 1983, 7 December 1984 and 27 June 1985 for a total of 9.8 ks, 9.3 ks
and 6.3 ks, respectively. 

\begin{figure}
\center
\begin{tabular}{c}
\resizebox{0.7\hsize}{!}{\includegraphics{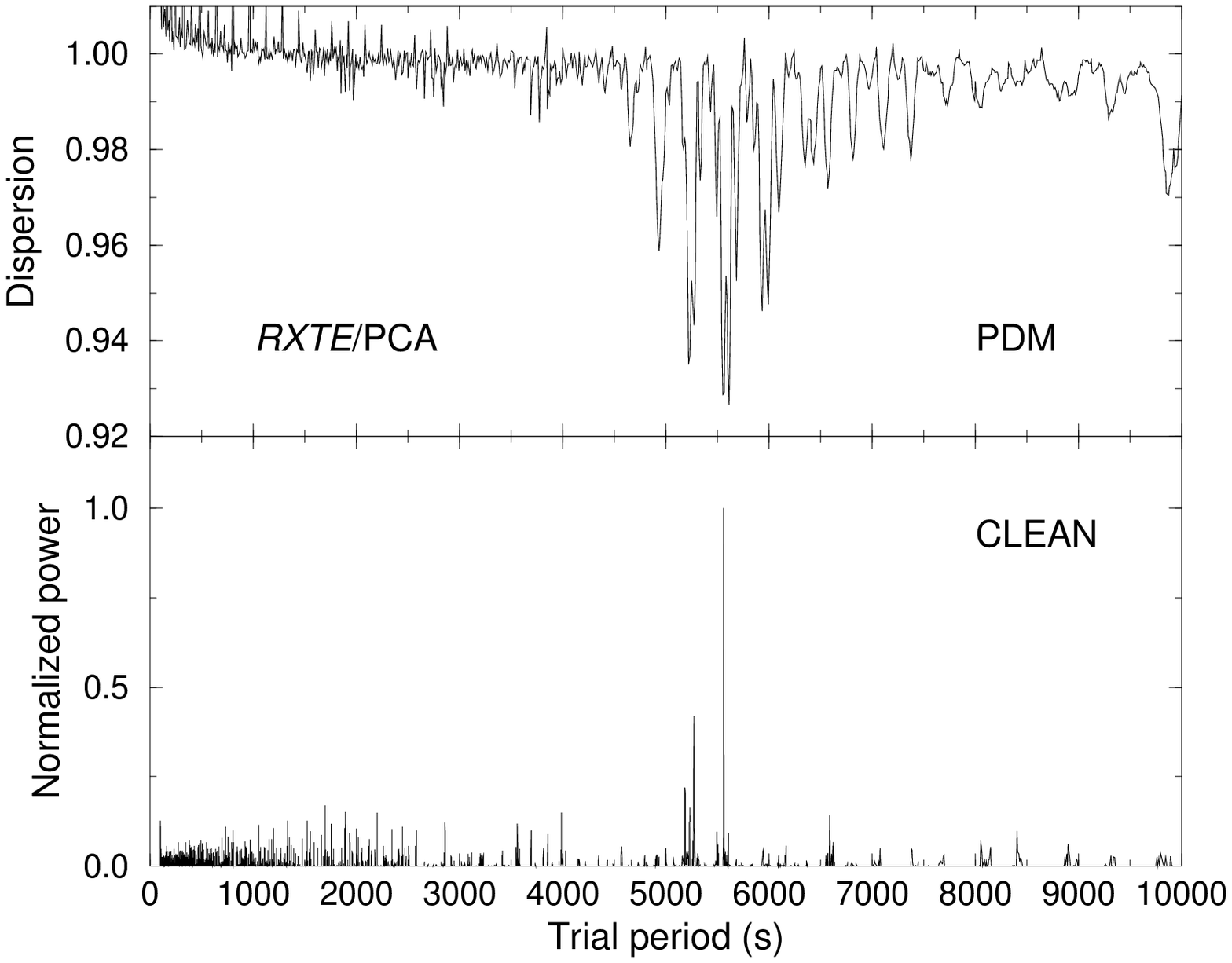}} \\
\resizebox{0.7\hsize}{!}{\includegraphics{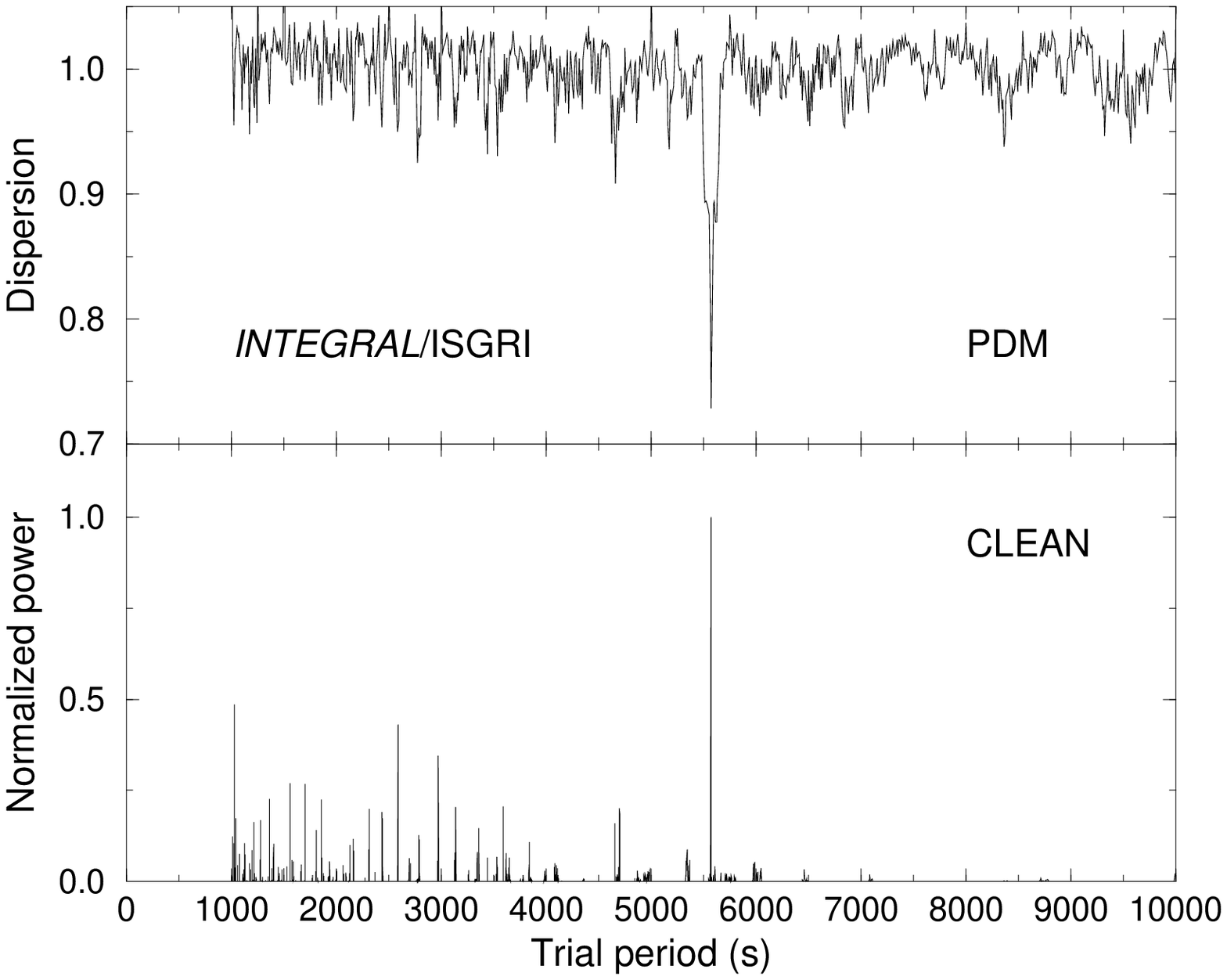}}\\
\resizebox{0.7\hsize}{!}{\includegraphics{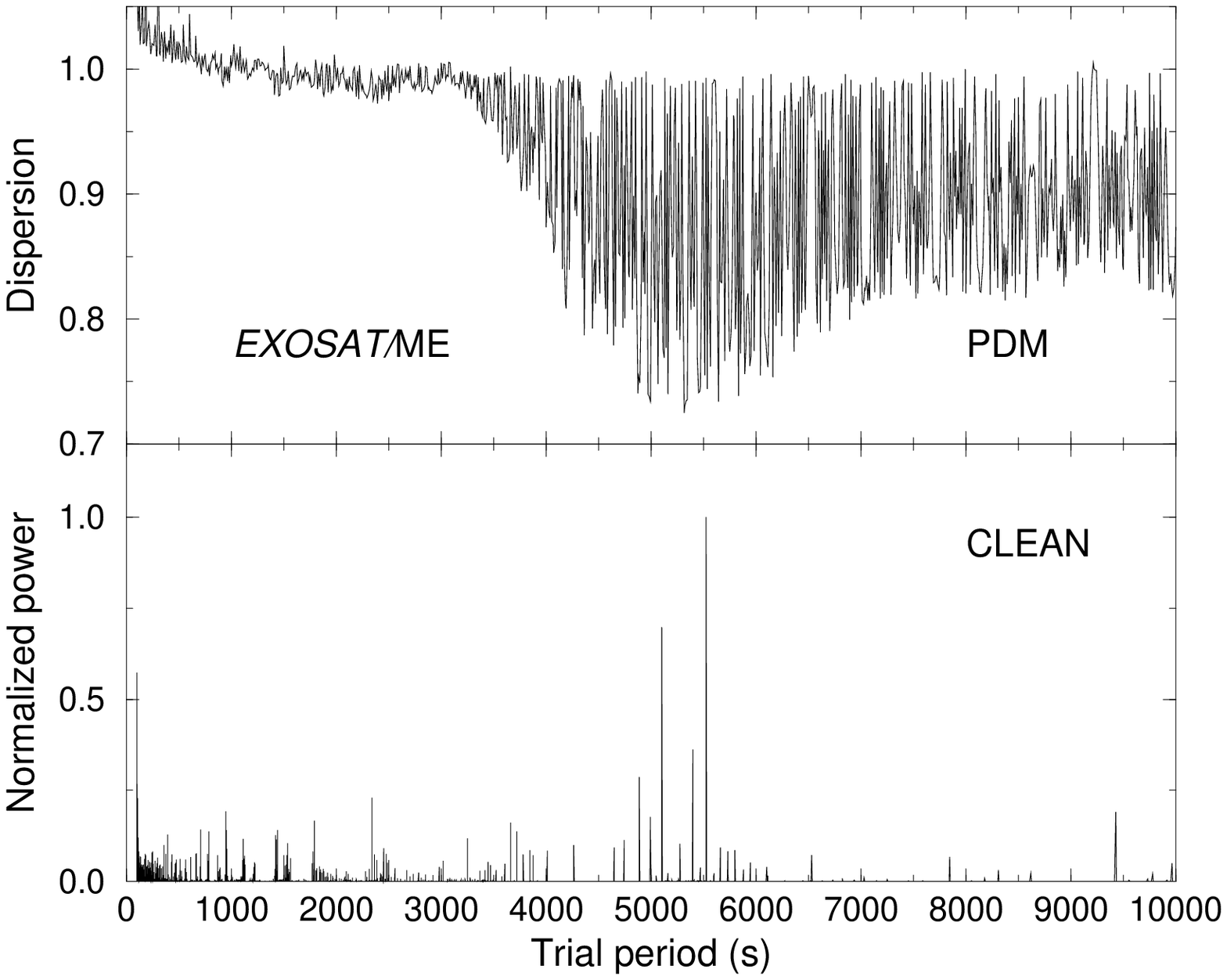}}\\
\end{tabular}
\caption{Periodograms of the entire light curves of \src\ for three
different instruments obtained by using the 
PDM and CLEAN
algorithms. A periodic signal is clearly detected at $\sim$5560~s.}
\label{pdm_clean}
\end{figure}
\begin{figure}
\center
\resizebox{1.0\hsize}{!}{\includegraphics{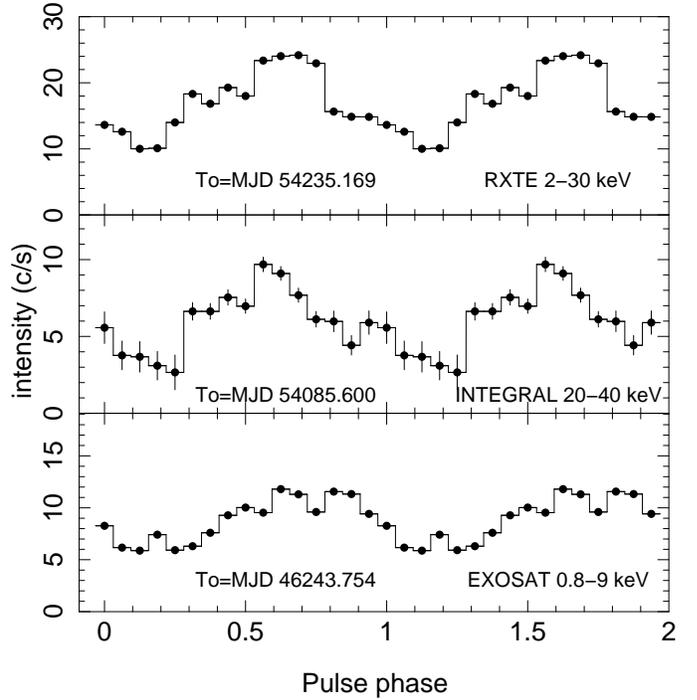}} 
\caption{Pulse profiles obtained from the light curves of {\it RXTE}/PCA, 
{\it INTEGRAL}/ISGRI, and {\it EXOSAT}/ME folded on to the period 
$P_{\rm pulse}$=5559~s. Two pulse periods are shown for clarity.}
\label{prof}
\end{figure}

\section{Timing analysis}

We have performed a timing analysis on two time scales: a) minutes
to hours to search for slow pulsations and b) days to investigate
orbital variability. Figure~\ref{lc} shows the entire {\it RXTE/PCA} light curve.
Each point represents the average of one observation interval (see
Table~\ref{xobs}).  On short time scales the X-ray light curve of \src\ is
dominated by erratic flaring, with changes in the X-ray intensity by a
factor 3--5 over a few minutes (Fig.~\ref{lc_int}). Despite the different
wind parameters and optical companion believed to be present in \src\
\citep{blay06,ribo06}, the light curves are very similar to those of X-ray
sources in which a neutron star is fed by direct accretion from the wind of
a supergiant, such as Vela X-1. This supports the idea that accretion in
this source is not mediated by an accretion disc, but proceeds directly
from the wind. Figure~\ref{lc_int} shows a portion of the light curve for
each interval. The $Y$-axis spans the same intensity range in all panels to
show the large variability of \src\ on short timescales. 

Figure \ref{hr} shows a colour-intensity diagram where three hardness
ratios are plotted as a function of 4--30 keV PCU2 count rate. Each point
represents an average over 512 seconds. The source becomes harder as the
count rate increases. A fit to a constant function does not give good fits
to the top two panels, with reduced $\chi^2_{\rm r}$ of 8.5 and 5.1,
indicating that the variation is statistically significant. In contrast,
the data of the bottom panel give $\chi^2_{\rm r}=1.4$.  The correlation
coefficients to a linear regression fits in the three panels are, from top
to bottom, 0.6, 0.7 and --0.2.

\subsection{Search for long-term pulsations}

In spite of intensive searches, X-ray pulsations have not been detected in
\src. Searches for X-ray coherent emission have been performed with {\it
EXOSAT} \citep[][but see \citealt{sara92}]{corb01}, {\it RXTE}
\citep{negu01,torr04,corb07}, {\it BeppoSAX} \citep{mase04,torr04} and {\it
INTEGRAL} \citep{blay05}. These studies demonstrate the lack of X-ray
pulsations on timescales from $\sim$ 1 ms to $\sim$ 1 hour. The possible
detection of 392-s pulsations reported by \citet{sara92} has never been
confirmed. \citet{corb01} reanalysed the {\it EXOSAT} observations used by
\citet{sara92} and did not find any evidence of any pulsations longer than
2 s. They argue that \citet{sara92} used a procedure not well suited to
search for coherent periodic pulses.

The only timescales that have not been investigated in detail due to either
technical constraints (low-Earth orbits) or lack of long uninterrupted
observations is from $\sim$ 1 hr to $\sim$ 1 day. The long coverage of our
{\it RXTE} observations allows us to investigate the higher frequency part
of this unexplored interval.

We searched for periodic signals in the light curves using various standard
techniques. Initially we applied an FFT to the two longest observation
intervals (see Table~\ref{xobs}). However, in order to be able to use the
entire data set, techniques suitable for handling missing data (gaps) in the
light curves are needed. Thus we employed PERIOD04 \citep{lenz05}, the
Phase Dispersion Minimisation (PDM, \citealt{stel78}) and the CLEAN
algorithm \citep{robe87}. A periodic signal around $\sim$5560~s is clearly
detected with all methods. The PDM and CLEAN periodograms obtained between
100 and 10\,000~s for the {\it RXTE}/PCA light curve binned at 16~s
intervals are shown in the top panel of Fig.~\ref{pdm_clean}.   In the case
of PDM there is a strong beating with $\sim$400~s separation, and two
equally significant minima occur at $5555\pm10$~s and $5608\pm10$~s. In the
case of CLEAN there is a clear maximum only at $5559\pm3$~s.

To further study the periodic signal detected with the whole data set, we
have analysed the {\it RXTE}/PCA light curves of intervals A+B and
intervals C+D+E separately, corresponding to a high and low flux states of
the source on an orbital timescale (Fig.~\ref{lc}). The results show that
the periodic signal is detected in both data sets, clearly indicating that
it is independent of the X-ray flux of the source. In particular, for A+B
PDM and CLEAN provide signals at $5550\pm50$~s and $5562\pm3$~s,
respectively. For C+D+E we obtain $5560\pm30$~s and $5559\pm3$~s,
respectively. The probability that the detected period is due to
random fluctuations is less than $10^{-8}$, according to the statistical
properties of the PDM algorithm \citep{stel78}.


We note that a strong signal is also detected around
815\,000$\pm$15\,000~s ($9.4\pm0.2$~d) with the CLEAN algorithm in the {\it
RXTE}/PCA light curve. Although the available database is limited to just
7.0~d, it is interesting to note that a clear hint of the possible 9.6
d (or of a part of the 19.25 d) orbital period appears in the {\it
RXTE}/PCA data set of \src. The probability that there is not such
periodicity or that the real period is different to 815\,000 s is less than
0.01.

The value of 5560 s is very close to {\it RXTE}'s orbital period --
{\it RXTE} follows a low-Earth circular orbit at an altitude of 580 km,
corresponding to an orbital period of about 90 minutes. Note that the {\it
RXTE}'s orbital period is continuously changing due to  perturbations by
the Earth and Moon and atmospheric drag. At the time of the observations
the orbital period laid in the range 5667.3--5669.1 s. In order to rule out
an instrumental origin of the modulation, we analysed archived data from
the {\it INTEGRAL}/ISGRI and {\it EXOSAT}/ME missions and instruments. The
orbit of these two observatories is highly eccentric with a period of 72
hours and 90 hours, respectively. The time bin was 500 s for the ISGRI
light curve and 30 s for the ME light curve.

The timing analysis on the {\it INTEGRAL} (20--40 keV) and {\it EXOSAT}
(0.8--9 keV) light curves gives periodic signals at $5570\pm20$~s and
$5525\pm30$~s, respectively (Fig.~\ref{pdm_clean}), which clearly confirms
the presence of the 1.54-hr periodic signal found with the {\it RXTE}/PCA
data.

Figure~\ref{prof} shows the pulse profile corresponding to each satellite.
The pulsed fraction is $\sim$50\% for the {\it RXTE} and {\it INTEGRAL}
light curves and 35\% for the {\it EXOSAT} data.

HEXTE data were not used in the periodicity analysis because of the
problems with  cluster A (see Sect. 2) and the low S/N of the cluster B
light curves. Also, note that the  energy range covered by HEXTE is
partially covered by ISGRI/{\it INTEGRAL}. 

\begin{table*}
\begin{center}
\caption{Results of the power-law fits to the power spectra.}
\label{psdfit}
\begin{tabular}{cccccccc}
\hline \hline \noalign{\smallskip}
	& MJD			&$I_{\rm X}^a$&Orbital&Number	&$\Gamma$	&rms (\%)&$\chi^2_{\rm r}$/dof 	\\
	&			&c/s	&phase$^b$&of PSD	&	&(0.001-1 Hz)& \\
\hline 
A	&54235.16--54236.11	&29	&0.87/0.93	&21	&1.87$\pm$0.04	&37	&1.3/44\\
B	&54236.86--54238.00	&27	&0.06/0.03	&26	&1.77$\pm$0.04	&34	&1.1/44\\
C	&54238.00--54239.06	&12	&0.17/0.09	&19	&1.55$\pm$0.05	&30	&1.9/43\\
D	&54239.09--54241.03	&6	&0.33/0.17	&24	&1.5$^{+0.09}_{-0.13}$	&28	&1.3/44\\
E	&54241.05--54242.20	&9	&0.50/0.25	&15	&1.5$\pm$0.1	&23	&1.2/44\\


\hline \noalign{\smallskip}
\multicolumn{8}{l}{$a$: Background subtracted 2--60 keV, PCU2 only}\\
\multicolumn{8}{l}{$b$: At mid-point for $P_{\rm orb}=9.56$ d and $P_{\rm
orb}=19.12$ d, respectively}\\
\end{tabular}
\end{center}
\end{table*}

\begin{figure}
\resizebox{\hsize}{!}{\includegraphics{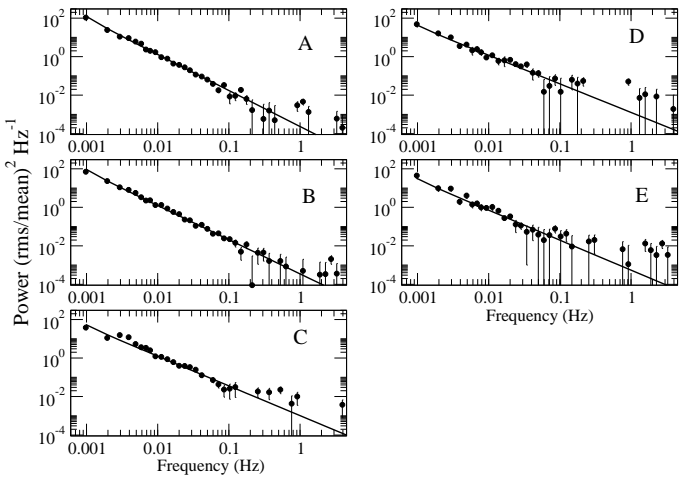}} 
\caption[]{Power spectra of \src\ at different instances of the
observations. See Table~\ref{psdfit} for a definition of the intervals.
}
\label{psd}
\end{figure}

\subsection{Orbital variability}

To investigate the longer-term variability, the observations were divided
into five intervals with roughly the same on-source time. The average
intensity decreased throughout the time covered by the observations as can
be seen in Fig.~\ref{lc}. The numbers on top of this figure are the orbital
phase assuming $P_1=9.56$ d and twice that period $P_2=19.12$ d
\citep{corb07}. The reference time is the epoch of maximum flux as given by
the ASM, MJD 53567.74 \citep{corb07}. The PCA light curve shows maximum
flux at phase $\sim$0, in agreement with the modulation of the folded ASM
light curve \citep{corb07}. There is another less significant peak at about
phase 0.5, if $P_{\rm orb}=9.56$ d.  If the orbital period were $P_{\rm
orb}=19.12$ d, then we would expect to see the maximum flux around phase
0.125 \citep{corb07}, which is not seen.



In order to investigate the evolution of the power spectral parameters of
the noise components as the source moves around the orbit, we produced one
power spectrum for each interval. The power spectra were obtained by
dividing the 2--20 keV PCA $2^{-7}$-s binned light curve of the entire
observation into 1024-s segments and an FFT obtained for each segment. The
contribution by the photon counting noise was computed and subtracted from
each power spectrum. The power spectra were normalized such that their
integral gives the squared fractional rms variability
\citep{bell90,miya91}. The power spectra of the same interval were averaged
to produce the final power spectrum. A logarithmic rebinning was also
applied to reduce the noise at high frequencies. Figure~\ref{psd} shows the
power spectra for the five intervals. The results of the fitting
procedure are shown in Table~\ref{psdfit}.

A single power-law component provides good fits for all intervals. The
power-law index and the fractional $rms$ amplitude  change with X-ray
intensity. The high-intensity intervals (A and B) show a steeper power
spectrum and a larger amplitude of variability. The low-intensity
intervals (D and E) display a flatter power law and are  less variable.
Interval C shows intermediate values between these two cases.

   \begin{figure}
   \centering
\resizebox{\columnwidth}{!}{\includegraphics[angle=-90]{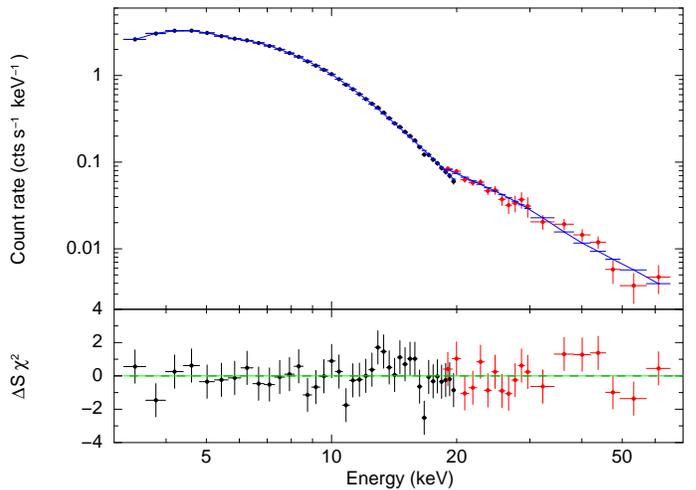}}
   \caption{X-ray spectrum in the energy range 3--70 keV obtained from PCA and 
   HEXTE and fitted with a 
   thermal Comptonization model plus a black body at low energies 
   (continuous line). Data to model residuals, in units of sigmas, 
   are depicted in the lower panel.}
              \label{fig:comptt}
    \end{figure}

\section{Spectral analysis}

In this section we present the results of the energy spectral
analysis. We first present the PCA+HEXTE time-average spectrum for the
entire observation. Then, we search for orbital variability in
the spectral parameters. 

\subsection{Time-averaged spectrum}

The 3-30 keV spectrum of \src\  is well described by  an absorbed, $N_{\rm
H}=1.3\pm0.5 \times 10^{22}$ cm$^{-2}$, power law with photon index
1.3$\pm$0.1 modified by a cutoff at 6.3$\pm$0.4 keV and folding energy
14$\pm$2 keV, in agreement with those found by \citet{negu01}. However, as
shown in \citet{torr04}, this model does not describe properly the spectrum
when data beyond 30 keV are taken into account ($\chi^{2}_{\nu}$=1.44 for
53 d.o.f.). The joined time-average  PCA+HEXTE spectrum
(Fig.~\ref{fig:comptt}) is, overall, well described by thermal
Comptonization (\texttt{comptt}  in \textsc{xspec} notation). This model
requires further a blackbody emission (\texttt{bb}) component to account
for an excess at low energies. The same is true when dynamical
Comptonization (\texttt{bmc}) replaces thermal Comptonization.  The
parameters for these two composite models are presented in Table
\ref{tab:compton_fits}. $kT_{\rm bb}$ is the temperature of the blackbody
component that accounts for the soft excess; $kT_{\rm 0}$ and $kT_{\rm
col}$ are the temperatures of the seed photons, which are subsequently 
Comptonized by the high temperature plasma, in the thermal and dynamical
Comptonization models respectively; $kT_{\rm e}$ and $\tau$ are the
temperature and optical depth of the high temperature plasma, in the
thermal Comptonization model; $\alpha$ is an spectral index related to the
efficiency of the Comptonization process (the lower the value of
$\alpha$, the more efficient is the Comptonization); $f$ is the
\emph{illumination factor}, i.e., the ratio of the number of multiply
scattered photons to the total number of seed photons ($f \gg 1$ implies
that the spectrum is fully Comptonized). No iron line is required. If the
absorption column is let free to vary, then the best-fit value is higher
than in previous observations. Fixing it to a lower value  increases
slightly the temperature of the blackbody but it is always around 0.5 keV
(for example fixing $N_{\rm H}$ to $0.3 \times 10^{22}$ cm$^{-2}$, the pure
interstellar value, gives $kT_{\rm bb}=0.56$ and sets the
$\chi^{2}_{\nu}=1$). Since {\it RXTE} cannot constrain the column density
due to the lack of sensitiveness below 3 keV we fixed $N_{\rm H}$ to the
value reported from {\it BeppoSAX} observations, namely $N_{\rm H}=0.9
\times 10^{22}$ cm$^{-2}$ \citep{mase04}. Applying Eq.~1 in \citet{torr04}
to the data in Table~\ref{tab:compton_fits}, we determine the radius of the
soft photon source which would produce the observed
luminosity\footnote{This formula gives the proper factor for a
\emph{circular} spot of area $\pi R_{W}^{2}$. Of course, a \emph{sphere}
would require a radius $R^{\rm sph}= R^{\rm spot}/2$ to produce the same
$L_{X}$ assuming isotropic emission.} to be $R_{W}=1.5$ km. This is only
consistent with a hot spot on the neutron star surface.

The possible cyclotron line around 30 keV reported by various authors
\citep{torr04,mase04,blay05} is not seen in the present data set. We did
not find significant evidence for the iron emission line at $\sim$6.4 keV
either. We estimate an upper limit on the equivalent width of a narrow line
(FWHM=0.1 keV) around 6.4 keV of 25 eV at 90\% confidence level. The
average X-ray luminosity in the energy range 3--30 keV is $2 \times
10^{35}$ erg s$^{-1}$, assuming a distance of 2.6 kpc \citep{blay06}.

\begin{table}
\begin{center}
      \caption[]{Parameters for the Comptonization models. Error bars are 90\%
      confidence level.}
         \label{tab:compton_fits}
\begin{tabular}{lc|lc}        
  \hline\hline \noalign{\smallskip}
\multicolumn{2}{c}{\texttt bb + comptt} & \multicolumn{2}{c}{\texttt bb + bmc}  \\  
\hline\noalign{\smallskip}
$N_{\rm H}^{(a)}$ 	&  0.9 						&$N_{\rm H}^{(a)}$ &  0.9 \\
 			& 						&		  &	\\
$kT_{\rm bb}$ (keV) 	& 0.55$^{+0.08}_{-0.07}$ 			&$kT_{\rm bb}$ (keV) & 0.46$^{+0.08}_{-0.10}$ \\
Norm$_{\rm bb}$ 	& (4.65$^{+1.25}_{-0.49}$)$\times 10^{-4}$	&Norm$_{\rm bb}$ & (4.17$^{+5.72}_{-1.09}$)$\times 10^{-4}$ \\
$F^{(b)}_{\rm bb}$ 	& 0.08						& $F^{(b)}_{\rm bb}$ & 0.04 \\
 			&						&		&		 \\
$kT_{0}$ (keV) 		& 1.30$^{+0.07}_{-0.05}$ 			&$kT_{\rm col}$ (keV) & 1.30$^{+0.05}_{-0.05}$	\\
$kT_{\rm e}$ (keV) 	& 35.20$^{+1.25}_{-1.26}$			&$\alpha$ & 1.33$^{+0.04}_{-0.05}$	 \\
$\tau$ 			& 0.73$^{+0.03}_{-0.04}$ 			&$f$ 		& $\gg$ 1	\\
Norm$_{\rm comptt}$ 	& (6.53$^{+1.05}_{-0.48}$)$\times 10^{-4}$ 	&Norm$_{\rm bmc}$ & (1.68$^{+0.01}_{-0.01}$)$\times 10^{-3}$	\\
$F^{(b)}_{\rm comptt}$ 	& 2.49						&$F^{(b)}_{\rm bb}$ & 2.52	 \\
 			& 						&		&		\\
$\chi^{2}_{\nu}$ (d.o.f.) & 1.00 (52)					&$\chi^{2}_{\nu}$ (d.o.f.) & 0.94 (53)	 \\
\hline \noalign{\smallskip}
\multicolumn{4}{l}{$^{(a)}$ Hydrogen column density in units of 10$^{22}$ cm$^{-2}$ (fixed)}\\
\multicolumn{4}{l}{$^{(b)}$ 3--60 keV unabsorbed flux in units of 10$^{-10}$ erg s$^{-1}$ cm$^{-2}$}
\end{tabular}
\end{center}
\end{table}

\begin{table*}
\begin{center}
      \caption[]{Results of the exponentially cutoff power law fits to the 
      energy spectra. In the bottom table $N_{\rm H}$ was kept fixed.}
         \label{orbspec}
\begin{tabular}{ccccccccc }        
  \hline\hline \noalign{\smallskip}  
  Bin  &orbital phase$^{(a)}$ & $N_{\rm H}^{(b)}$ & $E_{\rm cut}$ & $E_{\rm fold}$ &  $\Gamma$ & Norm & $F_{3-30}^{(c)}$ & $\chi^{2}_{\nu}$ (d.o.f.) \\
  \hline
  A &0.87/0.93 & 1.54$^{+0.89}_{-0.83}$ & 6.34$^{+0.68}_{-0.51}$ & 13.34$^{+2,49}_{-1.95}$ & 1.29$^{+0.15}_{-0.16}$ & 0.028$^{+0.009}_{-0.028}$ & 3.33 & 0.65 (51)\\
  B &0.06/0.03 & 1.54$^{+0.59}_{-0.58}$ & 6.32$^{+0.47}_{-0.38}$ & 14.46$^{+1.93}_{-1.60}$ & 1.32$^{+0.10}_{-0.11}$ & 0.032$^{+0.007}_{-0.006}$ & 3.67 & 0.72 (51) \\
  C &0.17/0.09 & 0.63$^{+1.01}_{-0.63}$ & 6.23$^{+0.60}_{-0.43}$ & 12.03$^{+2.58}_{-1.55}$ & 1.28$^{+0.18}_{-0.13}$ & 0.012$^{+0.004}_{-0.002}$ & 1.32 & 0.72 (51) \\
  D &0.33/0.17 & 0.94$^{+1.01}_{-0.94}$ & 6.13$^{+0.63}_{-0.51}$ & 11.99$^{+2.78}_{-2.20}$ & 1.44$^{+0.18}_{-0.20}$ & 0.009$^{+0.004}_{-0.003}$ & 0.76 & 1.37 (51) \\
  E &0.50/0.25 & 0.95$^{+1.11}_{-0.95}$ & 6.24$^{+0.45}_{-0.63}$ & 12.24$^{+3.20}_{-2.01}$ & 1.44$^{+0.18}_{-0.17}$ & 0.010$^{+0.004}_{-0.003}$ & 0.82 & 0.63 (51) \\
 \hline \noalign{\smallskip}
  A &0.87/0.93 & 0.9 & 6.00$^{+0.26}_{-0.27}$ & 11.88$^{+0.61}_{-0.57}$ & 1.17$^{+0.04}_{-0.04}$ & 0.023$^{+0.001}_{-0.001}$ & 3.26 & 0.67 (52)\\
  B &0.06/0.03 & 0.9 & 5.99$^{+0.22}_{-0.21}$ & 12.83$^{+0.57}_{-0.53}$ & 1.21$^{+0.03}_{-0.03}$ & 0.026$^{+0.001}_{-0.001}$ & 3.60 & 0.77 (52) \\
  C &0.17/0.09 & 0.9 & 6.34$^{+0.33}_{-0.33}$ & 12.62$^{+0.93}_{-0.82}$ & 1.33$^{+0.04}_{-0.05}$ & 0.013$^{+0.001}_{-0.001}$ & 1.33 & 0.63 (52) \\
  D &0.33/0.17 & 0.9 & 6.12$^{+0.33}_{-0.32}$ & 11.87$^{+0.94}_{-0.82}$ & 1.43$^{+0.04}_{-0.05}$ & 0.009$^{+0.001}_{-0.001}$ & 0.76 & 0.71 (52) \\
  E &0.50/0.25 & 0.9 & 6.25$^{+0.33}_{-0.39}$ & 12.17$^{+1.09}_{-0.96}$ & 1.44$^{+0.04}_{-0.06}$ & 0.009$^{+0.002}_{-0.001}$ & 0.82 & 1.33 (52) \\
\hline \noalign{\smallskip}
\multicolumn{8}{l}{$(a)$: At mid-point for $P_{\rm orb}=9.56$ d and $2P_{\rm orb}=19.12$ d, respectively}\\
\multicolumn{8}{l}{$(b)$: Hydrogen column density in units of 10$^{22}$ cm$^{-2}$}\\
\multicolumn{8}{l}{$(c)$: Unabsorbed flux in units of 10$^{-10}$ erg s$^{-1}$ cm$^{-2}$} \\
\end{tabular}
\end{center}
\end{table*}

   \begin{figure}
   \centering
\resizebox{\columnwidth}{!}{\includegraphics{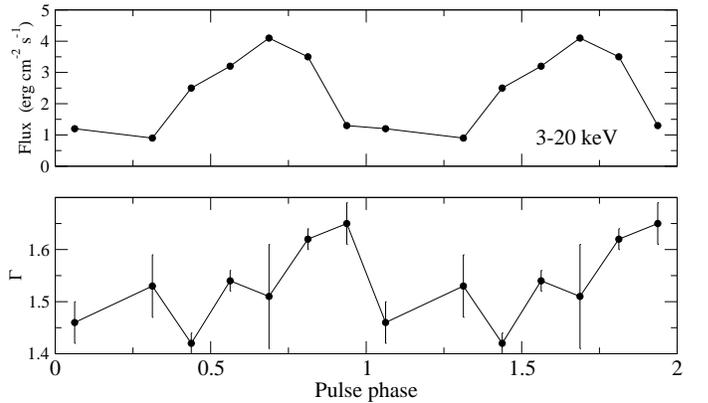}}
   \caption{X-ray flux and photon index as a function of the spin phase.}
              \label{pps}
    \end{figure}

\subsection{Pulse-phase spectroscopy}

In order to investigate how the X-ray spectrum changes as a function of the
spin of the magnetized neutron star, the pulse profile  corresponding to
intervals A and B was divided into eight equally spaced bins and a spectrum
in the energy range 3-20 keV was obtained for each bin. After
subtracting the underlying persistent emission, each spectrum was
satisfactorily fitted with an absorbed power law. The column density was
fixed to $1.3\times 10^{22}$ cm$^{-2}$, i.e., the value found in the
time-averaged spectrum (Sect. 4.1), although the use of $N_{\rm H}=0.9
\times 10^{22}$ cm$^{-2}$ did not change the results significantly.
Figure~\ref{pps} shows the variation of the 3-20 keV X-ray flux and the
power-law photon index as a function of the pulse phase. Phase zero
corresponds to MJD 54235.169. Marginal evidence for variability in the
power-law photon index is found. The spectrum becomes softer as the pulse
phase increases, with the photon index reaching a maximum around phase 0.9.
The hardness of the spectrum does not seem to depend on the pulse flux, but
the photon index appears to lag the pulse flux by about 1200 s. However,
the uncertainty in the photon index values is too large to draw a firm
conclusion.

\subsection{Orbital spectral variability}

To investigate possible orbital spectral variability, we obtained a
time-averaged spectrum for each observation  interval. The 3--30 keV band
spectrum was fitted to an absorbed power law modified at high energies by
an exponential cutoff. Given the high low-energy limit, the hydrogen column
$N_{\rm H}$ is not well constrained.  Although the observations that
correspond to a lower flux have lower column densities, the value of
$N_{\rm H}$ is consistent with no variability, within the errors. Thus, we
also tried fits with a fixed $N_{\rm H}=0.9\times10^{22}$ cm$^{-2}$ (lower
part of Table~\ref{orbspec}). In either case, a variation in the power-law
index is apparent. High-flux states show harder spectra.

\section{Discussion}

We have reported the analysis of the monitoring campaign of the high-mass
X-ray binary \src\ that took place in May 2007 with {\it RXTE}.  The
observations span over $\sim$ 7 days, which represent a large fraction
of the 9.6-day period. The long and uninterrupted observations allow us for
the first time to {\em i)} extend the search for X-ray pulsations to time
scales of about one hour, {\em ii)} investigate the X-ray orbital
variability and {\em iii)} give new insights into its energy emission
properties. In support of the results of these observations we also
analysed data from {\it INTEGRAL}/ISGRI and {\it EXOSAT}/ME.


\subsection{X-ray pulsations}

The proximity of \src\ to {\it RXTE}'s Continuous Viewing Zone allowed the
acquisition of long and uninterrupted observations, and the discovery of
slow X-ray pulsations ($P_{\rm spin}=5559\pm5$ s). With the discovery of
X-ray pulsations we have solved one of the questions that remained
unanswered, namely, the nature of the compact companion. The vast majority
of HMXBs harbour X-ray pulsars. These are
believed to be young neutron stars with relatively  strong magnetic fields
($B\sim10^{12}\:{\rm G}$). Among the handful of HMXBs not displaying X-ray
pulsations, three show the typical characteristics of persistent black-hole
systems. Of these, two are in the LMC and only one (Cyg X-1) in the
Galaxy.  The nature of the other three HMXBs in which pulsations have not
been discovered in spite of intensive searches  is under debate, although
two of them may also contain black hole companions: LS~I~+61~303/2E
0236.6+6101 \citep[][but see \citealt{dhaw06}]{casa05a} and
LS~5039/RX~J1826.2--1450  \citep[][but see \citealt{ribo08}]{casa05b}. If
confirmed it would explain the lack of pulsations in these systems.
Geometric effects (very low inclination or a very small angle between the
spin and rotation axises) have been invoked to explain the lack of
pulsations in  4U~1700--37, which might contain a massive neutron star
\citep{clar02,abub04}.  The lack of pulsations was more difficult to
explain in \src, as previous observations indicated  that this source
likely hosts a canonical neutron star accreting from the wind of a
main-sequence early-type star \citep{torr04,mase04,blay05,blay06,ribo06}.

With the period detected here, \src\ is the third (if the nature of IGR
J16358--4726 as a HMXB is finally confirmed) slowest HMXB pulsar, after
2S~0114+650 and IGR J16358--4726, and the first with a main-sequence
companion.  2S~0114+650 is a 2.78-h pulsar \citep{finl94} accreting matter
from a supergiant B1 companion \citep{reig96}, while IGR J16358--4726 has
been classified as a sgB[e] \citep[but see \citealt{nesp08}]{chat08} with a
1.6-hr rotating neutron star \citep{pate04}.\footnote{The slowest X-ray
pulsar is 4U~1954+319 with $P_{\rm spin}=5$ hr \citep{matt06,corb08}.
However, this is a symbiotic X-ray binary with a neutron star component accreting
from a low-mass M-type giant \citep{mase06b}.} 

Numerical calculations suggest that there is no significant angular
momentum transfer onto the neutron star from the wind of supergiant systems
\citep{ruff99}. Even a lower-velocity wind like the one believed to be
present in \src\ \citep{ribo06} seems to be insufficient to exert a spin-up
torque onto the neutron star. Since the transfer of angular momentum during
the accretion phase is so inefficient, a spin period of $\sim$ 5500 s must
have been attained in a previous evolutionary phase. The neutron star's
spin evolution in a close binary system \citep{davi79,davi81,dai06} is
divided into three phases, of which the accretion phase is the more recent
one. The neutron star is born as a millisecond pulsar (pulsar phase) and
spin down ocurs by magnetic dipole radiation until $P_{\rm spin}\sim 1$ s.
Then the neutron star enters the propeller phase, in which the neutron star
spins down because accretion is centrifugally inhibited. This phase
continues until the spin period reaches the so-called equilibrium period
given by \citep[][but see \citealt{davi81}]{li99}

\[P_{\rm eq} \approx 20
B_{12}^{6/7}\dot{M}_{15}^{-3/7}R_6^{18/7}M_{1.4}^{-5/7} \,\,\, {\rm s} \]

\noindent where $B=10^{12}B_{12}$ G is the neutron star's dipolar magnetic
field strength, $\dot{M}=10^{15}\dot{M}_{15}$ g s$^{-1}$ is the mass
accretion rate, $R=10^6R_6$ cm is the radius of the neutron star, and
$M=1.4M_{1.4} \msun$ is the mass of the neutron star. Adopting the
canonical values  $R_6=M_{1.4}=1$ and $B_{12}=3.6$ \citep{blay05} and
$\dot{M}_{15}\approx L_{\rm X}R/GM=0.5$ for $L_{\rm X}=10^{35}$ erg
s$^{-1}$, the \src\ equilibrium period results $P_{\rm eq}\sim80$ s, far
lower than the observed period. We conclude then that the current spin
period of \src\ cannot be explained in evolutionary terms. A way out is
provided by \citet{li99}, who explain the long spin period of 2S\,0114+65
assuming that the neutron star was born as a magnetar, that is, with
$B\simmore 10^{14}$ G. Indeed, the propeller effect can spin down the
neutron star to $5 \times 10^3$ s keeping the same values as above but
taking $B_{12}$ two orders of magnitude higher.


\subsection{Temporal variability}

The range in X-ray luminosity (3--30 keV) implied from the observations is
0.6--3.1 $\times 10^{35}$ erg s$^{-1}$,  assuming a distance to the source
of 2.6 kpc \citep{blay06}. That is, \src\ displays an amplitude of
variability in X-ray flux of about 5. This amplitude is larger than the
factor of 2 reported by \citet{ribo06} from ASM data, i.e., for X-rays in
the 1.3--12 keV band, but consistent with the observations by \citet[][see
their Fig. 4]{corb01}. Note that the results from the ASM are highly
dependent on the minimum flux, which for the ASM is uncertain. In the
simplified Bondi-Hoyle accretion model a maximum to minimum orbital X-ray
flux ratio of 5 can be explained assuming that the orbital variability is
due to changes in the mass accretion rate. In this scenario, an
eccentricity of $\sim 0.4$ is needed, somehow higher than that suggested
by \citet{ribo06}.

In addition to the X-ray flux, the power and energy spectral parameters
also display variability on time scales of days. If the orbital period of
9.6 days is assumed to be correct, then it seems natural to attribute this
variability to the motion of the source through the orbit. Note however,
that our results are based on less than one orbit. The power spectra of
\src\ are well described by a single power law, whose index and rms
amplitude  decrease as the count rate decreases (Table~\ref{psdfit}). 
There is no sign of a break at low frequencies (flat-topped noise) in the
frequency range covered by the power spectra. Red noise dominates the power
spectrum down to frequencies as low as 1 mHz.  The decrease in the
amplitude of the X-ray variability (rms) with decreasing  X-ray flux
seems to indicate a large scale change in the structure of the wind.  One
possible explanation is that the wind was highly inhomogeneous during 
intervals A, B and C, during periastron passage, as usually observed in 
systems with supergiant companions, and became smoother further out, in 
intervals C and D.  Assuming a mass of 18$M_{\odot}$ for the O9.5V primary
and the canonical neutron star mass of 1.4$M_{\odot}$ for the
accretor, the change in the wind structure must have occured in  between
intervals C and D, which in terms of radial distance corresponds to
$\sim$5.7 $R_*$ for $P_1=9.6$ d and $\sim$6.95 $R_*$ for $P_2=19.25$ d. The
detection of X-ray emission in intervals D and E in \src\ contrasts with
what is observed in classical supergiant systems for which the X-ray
emission is hardly detected when the compact object is located beyond 3
$R_*$ \citep{negu08,blay08}.

The 3--30 keV spectrum of \src\ can be described with a power law with
$\Gamma=1.5$ and a high-energy cutoff with $E_{\rm cut}=6.7$ keV (see Sect.
4.1). The power-law index increases as the X-ray flux decreases
(Table~\ref{orbspec}). The high-flux intervals (A and B) show a harder
spectrum than the low-flux states (C, D and E). This behaviour is the
opposite to that found in other X-ray sources. The explanation could be,
again, in the low X-ray brightness of this source. Within the framework of
bulk motion Comptonization, that is, assuming that Comptonization occurs in
the accretion flow near the neutron star surface (accretion column), the
surface of the neutron star acts as source of feedback against the infalling
material: when the accretion increases, so does the X-ray flux. The ram
pressure exerted by the X-ray source tends to hamper accretion and
therefore complicates the Comptonization process. In the very bright X-ray
sources, the feedback decreases the efficiency of Comptonization and
produces a high-soft low-hard behaviour. In \src\ however, with its main
sequence donor, the X-ray source is never pushed to the limit where the
X-rays can hamper the accretion process. With no feedback from the surface
of the neutron star, the source becomes harder when accreting material
becomes available, presumably at the periastron passage.

One of the controversial results of \src\ in the X-ray band is the shape of
the hardness-intensity diagram. \citet{negu01} found a correlation between
the X-ray colours and the X-ray intensity, in that the X-ray spectrum
becomes harder as the flux increases. Figure~\ref{hr} confirms these
results. This figure shows the hardness-intensity diagram for three
different ratios. In all three cases the change in intensity is
approximately the same, namely, a factor of 10. In contrast, \citet{mase04}
did not find any variability of the hardness ratio with intensity in a {\it
BeppoSAX} observation on 1998 November. \citet{mase04} suggested that the
hardness-intensity dependence might become evident at higher luminosities,
as the X-ray luminosity during the \citet{negu01} observations was about
one order of magnitude higher than that of \citet{mase04}. To investigate
this possibility we obtained hardness-intensity diagrams for interval B and
D, that is, when the X-ray flux was maximum and minimum, respectively. We
define the hardness ratios as $HR1=$7--10~keV/4--7~keV and 
$HR2=$10--15~keV/4--7~keV. Fitting $HR1$ and $HR2$ as a function of 4--30
keV intensity to a constant does not give acceptable fits for interval B.
The fits to interval D are acceptable (in the sense $\chi^2_{\rm r} < 2$).
However, the reason for this acceptable fit is because the error bars are
larger. As an example, a linear fit ($y=ax+b$) to the $HR2$-intensity
diagram of interval D reduces the $\chi^2$ from 419 (259 dof) to 290 (258
dof). An F-distribution test gives a chance improvement of only
0.18\%. That is, the addition of a linear component is significant above a
3-$\sigma$ level. In fact, irrespective of the X-ray flux, as the intensity
increases by a factor 10, the hardness ratios $HR1$ and $HR2$ increase by
$\sim$25\%. In summary, the hardness-intensity dependence becomes more
evident at higher source luminosities (as \citet{mase04} had suggested).
However, this effect is most likely due to low statistics rather than the
disappearance of the correlation. 

   \begin{figure}
   \centering
\resizebox{\columnwidth}{!}{\includegraphics{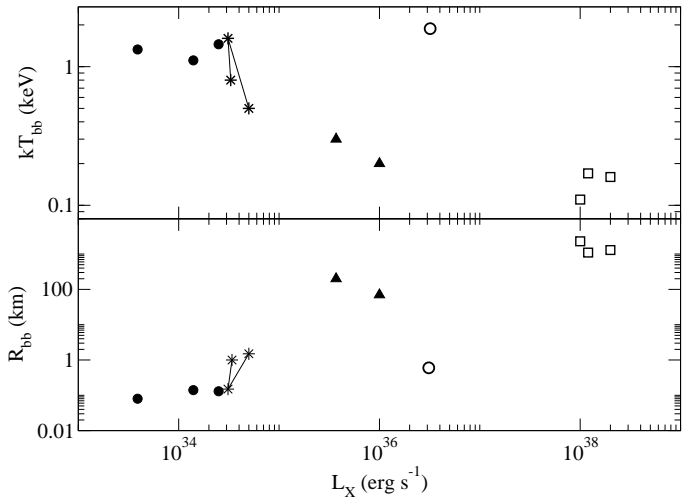}}
   \caption{Variation of the characteristic temperature (upper panel) and 
   radius (lower panel) with X-ray luminosity of the
   emitting region producing the soft excess, if modelled with a blackbody.
   Squares represent high-luminosity disc-fed supergiants (Cen X-3, SMC X-1
   and LMC X-4); filled circles are
   low-luminosity persistent Be/X-ray binaries (X-Per, A\,0535+262 and LS I
   +61 235); triangles correspond to
   the wind-fed supergiants (Vela X-1 and 2S 0114+65); the open circle is
   SAX J2103.5+4545; stars mark the 
   position of \src\ for three different observations (see text for
   details).}
   \label{soft}
    \end{figure}

\subsection{Soft component}

When a broader energy range is considered, the high-energy cutoff
power-law model does not give acceptable fits and a soft excess at energies
below 2 keV shows up in the residuals. More physical models involve
Comptonization. Either thermal or bulk-motion Comptonization produce good
fits to the X-ray energy continuum in the 3--80 keV band. In either case a
blackbody component with $kT_{\rm bb}=0.5$ keV is required to account for the
low-energy photons. 

The presence of a blackbody component supports the suggestion by 
\citet{hick04} that soft excess is common (probably ubiquitous) in X-ray
pulsars. For high luminosity ($L_{X}\sim 10^{38}$ erg s$^{-1}$) pulsars,
the soft component stems from reprocessing of hard X-rays from the
neutron star in the inner edge of an accretion disc. In less luminous
systems, other mechanisms, such as emission from diffuse gas or from the
neutron star surface, can be at work. To get a rough estimate of the size
of the soft radiation emission site, we compute the radius of the circular
blackbody spot that would produce the observed soft luminosity, according
to data in Table \ref{tab:compton_fits}. We get a radius of $R_{\rm bb}\sim
1.5$ km. This value is too small to come from the entire surface of the
neutron star, a hot corona, or an accretion disc. The lack of a disc in
\src\ was discussed in \citet{torr04}. This size is of the same order as
that deduced previously from the temperature of the seed soft photons
$R_{\rm W}\approx 1.5$ km. Then we tried to tie the temperature of both
components during the fits. The resulting statistics were significantly
worse. It seems, therefore, that the spectrum comes from two sites, with
the same characteristic size but different temperature. They could reflect
a gradient in temperature from the hot spot (of radius $R_{\rm W}$ in the
base of the accretion column) and the surface surrounding the spot (just
outside the column) and into the relatively cold surface.

When the soft excess is modelled with a blackbody component, the blackbody
spectral parameters (temperature and emission radius) strongly depend on
the source luminosity, which in turn, depends on the accretion mechanism.
In Fig.~\ref{soft}, we plot the temperature and radius of the emitting
region allegedly causing the soft excess. High-luminosity systems display
lower temperatures and larger radii (open squares). In high-luminosity
pulsars, such as SMC X-1, Cen X-3 and LMC X-4, accretion occurs via Roche
lobe overflow and an accretion disc is formed. In these systems $kT=0.1$
keV and $R_{\rm bb}$=100 km, typically \citep[see e.g. Table 5
in][]{hick04}, and the soft excess is believed to be due to reprocessing of
hard X-rays by optically thick accreting material. In contrast,
low-luminosity systems (filled circles), such as X Per \citep{cobu01}, A
0535+26 \citep{mukh05} and RX J0146.9+6121 \citep{palo06} have $kT=1.2$ keV
and $R_{\rm bb}=0.1$ km. In these systems the neutron star captures
material from the high-density wind from the circumstellar disc of the Be
companion. The origin of the soft excess would be (part of) the neutron
star surface. Note also that the high-luminosity pulsars mentioned above
contain supergiant companions, while the low-luminosity systems harbour
main-sequence stars. 
In between these extreme cases  one finds the wind-fed supergiant systems
(filled triangles), such as Vela X-1, with $kT_{\rm bb}=0.2$ keV and
$R_{\rm bb}=70$ km \citep{hick04} and 2S 0114+650, with $kT_{\rm bb}=0.3$
keV and $R_{\rm bb}=200$ km \citep{mase06a}. In these systems the thermal
component responsible for the soft excess is likely to be a cloud of
diffuse plasma around the neutron star \citep{hick04}. 

The star-like symbols in Fig.~\ref{soft} represent three different
observations of \src\ \citep[][and present work]{mase04,torr04}. \src\
appears close to the Be/X-ray binaries, as expected if the soft excess
arises from a hot spot on the neutron star surface. It is not clear how
such a hot spot can be formed in \src\ as accretion is also driven through
a stellar wind \citep{ribo06}. Perhaps the much slower (and presumable
denser) wind in \src\ might make the accretion configuration be more
similar to that of persistent Be/X-ray binaries. 

The open circle corresponds to the peculiar Be/X-ray binary SAX
J2103.5+4545 \citep{inam04}. This source is similar to \src\ in that the
spin and orbital periods are typical of supergiant X-ray binaries, but the
primary component of the binary is a main-sequence star. In both systems
the orbits have  $e=0.4$. The X-ray variability is, however, rather
different. SAX J2103.5+4545 presents X-ray bright states that last for a
few months and extended (few years) X-ray faint states. During the bright
states an accretion disc is formed around the neutron star \citep{bayk02}
and a decretion circumstellar disc is formed around the Be star
\citep{reig04}. These discs are probably short lived and appear during the
high X-ray emission states only. \citet{reig05} found that SAX J2103.5+4545
was emitting X-rays even after the complete loss of the circumstellar disc.
In this state, \citet{reig05} argue that the X-rays are the result of
wind-fed accretion, that is, as in \src.

\begin{figure}
   \centering
\resizebox{\columnwidth}{!}{\includegraphics{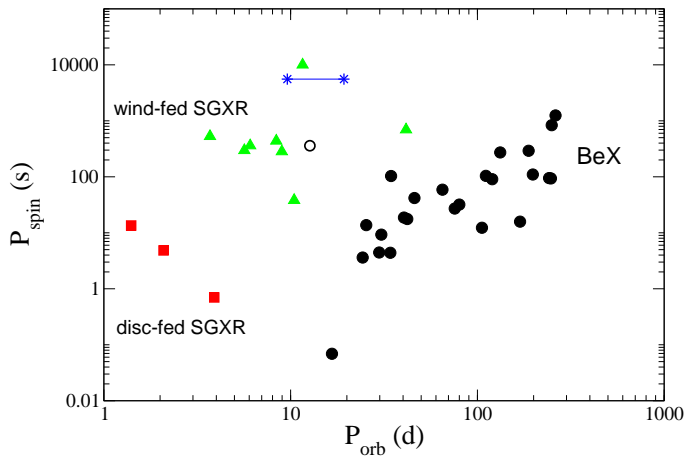}}
   \caption{$P_{\rm spin}$-$P_{orb}$ diagram. The two joined star-like
   symbols represent \src\ for the two suggested orbital periods. The open
   circle corresponds to SAX J2103.5+4545.}
   \label{pspo}
    \end{figure}

The position of these two sources in the  $P_{\rm spin}$-$P_{orb}$ diagram
(Fig.~\ref{pspo}) reinforces their similarity. They fall in the wind-fed
supergiant region. Since this diagram reflects the type of 
mass-loss/accretion mechanism \citep{corb86}, it is not surprising to find
\src\ and SAX J2103.5+4545 in this region. What makes them peculiar is
that they both contain a main-sequence donor.

The blackbody temperature and radius (scaled to a distance of 6.5 kpc;
\citealt{reig04}) of SAX J2103.5+454 shown in Fig.~\ref{soft} (open circle)
were obtained during a bright state \citep{inam04}, which may explain the
isolated position in the plot. Note that the value of the radius and
blackbody temperature are consistent with those of the low-luminosity
Be/X-ray binaries (filled circles). The difference appears in the X-ray
luminosity. The higher luminosity in SAX J2103.5+454 can be explained if an
accretion disc was present during the observations. The detection of a
quasi-periodic oscillation at 0.044 Hz \citep{inam04} and the correlation
between spin-up/down with X-ray flux \citep{bayk07} indeed supports
the presence of an accretion disc. 

\section{Conclusion}

We have finally unveiled the nature of the compact companion in the
high-mass X-ray binary \src. The {\it RXTE} light curve of \src\ appears to
be modulated with a period of $\sim$1.54 hours that we interpret as the
spin period of the neutron star. Evidence for this modulation is also found
in {\it INTEGRAL} and {\it EXOSAT} data. Thus we rule out that the
periodicity is due to the {\it RXTE} orbit. \src\ becomes the  slowest
rotating neutron star in a high-mass X-ray pulsar with a main-sequence
companion. The X-ray flux and the spectral index of the power-law
components that fit  energy and power spectra exhibit variability on time
scales of days. The amplitude of variability of the X-ray flux changed by a
factor of 5 on time scales of days. The power-law index is larger (i.e.
steeper spectrum) during low-flux states in the energy spectra and during
high-flux states in the power spectra. We attribute this variability to
changes in the mass accretion rate as the neutron star orbits the O-type
companion in a moderately eccentric orbit.  The time span covered by the
observations is shorter than any of the two suggested orbital periods,
hence further monitoring campaigns covering a larger fraction of the orbit
are needed to pin down the correct orbital period. However, the fact that
the maximum of the PCA light curve occurs at phase zero, i.e., in
coincidence with the ASM maximum when folded onto the 9.6-day period and
the detection of a strong signal at $\sim$815000 s in the {\it RXTE}
periodogram would favour the shorter period. The detection of a soft
component in the energy spectrum has been interpreted as emission from the
polar caps. The plot of the blackbody emitting radius and temperature as a
function of X-ray luminosity constitutes a useful tool to distinguish the
type of accretion mechanism at work in the X-ray binaries that show a soft
excess.



\begin{acknowledgements}

This research was supported by the European Union Marie Curie grant
MTKD-CT-2006-039965. This research is supported by the DGI of the Spanish
Ministerio de Educaci\'on y Ciencia under grants AYA2005-00095 and
AYA2007-68034-C03-01 and FEDER funds. JMT acknowledges the support of the
Spanish Ministerio de Educaci\'on y Ciencia  under grant PR2007-0176.  This
research has made use of data obtained through the INTEGRAL Science Data
Center (ISDC), Versoix, Switzerland.


\end{acknowledgements}

\end{document}